# The Shape of FIREbox Galaxies and a Potential Tension with Low-mass Disks

Courtney Klein[1], Jenny D. Wang[1], Luke Y. Xia[1], James S. Bullock[1,2], Jorge Moreno[3,4], Robert Feldmann[5], Francisco J. Mercado[3], Claude-André Faucher-Giguère[6], Jonathan Stern[7], N. Nicole Sanchez[8,9], Abdelaziz Hussein[10], and Michelle S. Park[11]

[1] Department of Physics and Astronomy, University of California Irvine, Irvine, CA 92697, USA
[2] Department of Physics and Astronomy, University of Southern California, Los Angeles, CA 90089, USA
[3] Department of Physics and Astronomy, Pomona College, Claremont, CA 91711, USA
[4] Carnegie Observatories, Pasadena, CA 91101, USA
[5] Department of Astrophysics, Universität Zürich, Winterthurerstrasse 190, CH-8057 Zürich, Switzerland
[6] CIERA and Department of Physics and Astronomy, Northwestern University, Evanston, IL, USA
[7] School of Physics and Astronomy, Tel Aviv University, Tel Aviv 69978, Israel
[8] The Observatories of the Carnegie Institution for Science, 813 Santa Barbara Street, Pasadena, CA 91101, USA
[9] Cahill Center for Astronomy and Astrophysics, California Institute of Technology, MC249-17, Pasadena, CA 91125, USA
[10] Department of Physics and Kavli Institute for Astrophysics and Space Research, Massachusetts Institute of Technology, 77 Massachusetts Avenue, Cambridge, MA 02139, USA
[11] Institute for Computational and Mathematical Engineering, Stanford University, Stanford, CA 94305, USA

## Abstract

We study the intrinsic and observable shapes of approximately 700 star-forming galaxies with stellar masses of $10^8$–$10^{11} M_\odot$ from the FIREbox simulation at $z = 0$. We calculate intrinsic axis ratios using inertia tensors weighted by three morphology types: "All Stars," "Young Stars," and "Luminosity-weighted Stars." Young Stars shows mass-dependent 3D configurations, with spheroidal, elongated, and disky shapes dominant at stellar masses of $10^{8.5}$, $10^{9.5}$, and $10^{10.5}$ $M_\odot$, respectively. Using the radiative transfer code SKIRT, we construct mock images for each galaxy and show that projected short-to-long axis ratios, $q$, inferred from 2D Sérsic fits are most closely related to Luminosity-weighted Stars tensor shapes and least resemble the All Stars' shapes. This suggests observed 2D shape distributions should not be compared to predictions based on 3D stellar mass shapes. Next, we construct a sample of mock images projected in random orientations and compare them to observed axis ratio distributions from the GAMA survey. At stellar masses below $10^{10}$ $M_\odot$, we produce too few galaxies with observed $q < 0.4$ and none with $q < 0.2$, suggesting that FIREbox does not produce enough low-mass disk galaxies. At higher masses, $10^{10}$–$10^{11}$ $M_\odot$, we find that the predicted $q$ distribution is sensitive to the dust-to-metal ratio; using our fiducial model, the distribution of $q$ values is formally consistent with observations, but there is tension with our ability to produce enough very thin systems with $q < 0.2$. Future observational and theoretical programs aimed at understanding disk and thin-disk fractions will provide crucial tests of galaxy formation models.

## 1. Introduction

The morphology and structure of galaxies is a topic of historic significance in the field of galaxy evolution (E. P. Hubble 1926; A. Sandage et al. 1970; J. Binney 1985). Galaxy shapes encode information about their evolution and act as important tests for theories of galaxy formation (S. G. Carlsten et al. 2021; M. Haslbauer et al. 2022).

It is useful to characterize the 3D shape of a galaxy as a triaxial ellipsoid with three principal axes: $A > B > C$ (J. Binney 1978). With this approximation, galaxy shapes may be classified into three categories: spheroidal (with three axes approximately equal), disky (oblate, with one axis much shorter than the other two), or elongated (prolate, with two axes much shorter than the long axis). Note that all three axes can be different, but we are still able characterize them as more triaxial oblate or prolate shapes as defined in A. van der Wel et al. (2014). Determining the 3D shape of any specific galaxy is difficult because, for any observed axis ratio observed, there could be a range of corresponding 3D axis ratios responsible. However, subject to priors, it is possible to statistically infer 3D shape distributions from observed 2D axis ratio distributions (G. Fasano & R. Vio 1993; B. S. Ryden 2004; B. P. Holden et al. 2012; Y.-Y. Chang et al. 2013; V. Pandya et al. 2024).

It is common to derive the 2D axis ratios of observed galaxies using best-fit Sérsic light profiles, often with packages such as GALFIT (C. Y. Peng et al. 2002), AstroPhot (C. J. Stone et al. 2023), or PySérsic (I. Pasha & T. B. Miller 2023). Alternatively, axis ratios can be determined through the ellipticity of a specific isophote (M. Franx et al. 1991; C. Stoughton et al. 2002; I. D. Karachentsev et al. 2013), the second moment of the galaxy image (C. Stoughton et al. 2002), or kinematics (M. Franx et al. 1991). Studies reliant on these methods have concluded that the majority of massive quenched galaxies in the local Universe are spheroidal and that massive star-forming galaxies at $z = 0$ are usually disks (D. G. Lambas et al. 1992; N. D. Padilla & M. A. Strauss 2008; S. de Nicola et al. 2022). Most studies suggest that low-mass galaxies at $z = 0$ are a mixture of thick disks and spheroids (S.-I. Ichikawa 1989; B. Binggeli & C. C. Popescu 1995; S. Roychowdhury et al. 2013; J. Putko et al. 2019; E. Kado-Fong et al. 2020; Y. Rong et al. 2020), although there are suggestions that a substantial population of local low-mass galaxies may be elongated in shape (R. A. Vincent & B. S. Ryden 2005; A. Burkert 2017).

Similar analyses using data from higher-redshift studies suggest that, in the early Universe, low-mass galaxies tend to be less disky and more elongated than seen at $z = 0$ (A. van der Wel et al. 2014; J. Zhang et al. 2022; V. Pandya et al. 2024), suggesting that small galaxies develop more rotational support over time. In contrast, the most massive galaxies in the early Universe tend to be more disky than those at late times, due to the evolution toward more quenched, spheroidal ellipticals as we approach $z = 0$ (B. P. Holden et al. 2012; Y.-Y. Chang et al. 2013; V. Pandya et al. 2024).

Compared with observations, modern cosmologically based numerical simulations have been successful in producing galaxies of varied shapes. Traditionally, 3D axis ratios of simulated galaxies have been calculated with star particles directly using 3D isodensity surfaces, mass eigentensors, or minimum-volume enclosing ellipsoids (e.g., D. Ceverino et al. 2015; M. Tomassetti et al. 2016; A. C. R. Thob et al. 2019; J. Zhang et al. 2022; B. Keith et al. 2025). However, since observations typically measure light profile shapes, projecting theoretically derived 3D shapes based on mass distributions will not necessarily provide a fair comparison to observed morphologies.

Forward modeling provides a way to overcome this difficulty (e.g., C. Bottrell et al. 2017; V. Rodriguez-Gomez et al. 2019; C. Bottrell et al. 2024). For example, A. De Graaff et al. (2022) used axis ratios determined from Sérsic fits of mock observations of EAGLE simulations and found a lack of thin, small axis ratio systems for their sample, which had $M_\star > 10^{10} M_\odot$. As another example, M. Haslbauer et al. (2022) used mass-derived 3D shapes of simulated Milky Way–mass galaxies from Illustris (D. Nelson et al. 2015), IllustrisTNG (D. Nelson et al. 2019), and EAGLE (J. Schaye et al. 2015) to construct predicted 2D shape distributions to compare to observations. They concluded that there were too few galaxies with small axis ratios in the simulations and suggested that modern galaxy formation models have difficulty producing enough thin-disk galaxies. However, D. Xu et al. (2024) analyzed the IllustrisTNG galaxies using mock images and concluded that the predicted distributions of Milky Way–mass galaxies were consistent with the observed axis ratio distributions.

In this paper, we measure the 3D and 2D shapes of $z = 0$ simulated galaxies in the FIREbox simulation (R. Feldmann et al. 2023). We measure their 3D axis ratios using a weighted inertia tensor applied in three different ways: (1) to all stars in each galaxy (the "All Stars" morphology), (2) to stars with ages < 0.5 Gyr (the "Young Stars" morphology), and (3) to $r$-band luminosity-weighted stars (the "Luminosity-weighted Stars" morphology). We then measure the 2D shapes of each galaxy by constructing mock images and compare the resultant shape distributions to results from the Galaxy and Mass Assembly (GAMA) survey. Section 2 discusses our simulations and the methods we use to determine shapes in 3D and 2D. Section 3.1 presents our 3D shapes and compares them to shapes inferred from idealized mock images in 2D. Section 3.2 compares our 2D shape distributions to observations. We discuss our results in Section 4.

## 2. Methods

### 2.1. Simulations

For this analysis we use FIREbox, a cosmological simulation with a volume of $(15 \text{ cMpc } h^{-1})^3$ from the Feedback in Realistic Environments (FIRE) project (R. Feldmann et al. 2023). FIRE-2 physics (P. F. Hopkins et al. 2018) was implemented in FIREbox using GIZMO with a meshless finite-mass Lagrangian Godunov method (P. F. Hopkins 2015). FIREbox initial conditions were established using the Multi-Scale Initial Conditions tool (O. Hahn & T. Abel 2011). A flat ΛCDM cosmology was assumed with $\Omega_m = 0.3089$, $\Omega_\Lambda = 1 - \Omega_m$, $\Omega_b = 0.0486$, $h = 0.6774$, $\sigma_8 = 0.8159$, and $n_s = 0.9667$ (Plank Collaboration et al. 2016).

FIREbox has a baryonic (dark matter) mass resolution of $6.3 \times 10^4 M_\odot$ ($3.3 \times 10^5 M_\odot$) with a star (dark matter) particle force softening length of 12 pc (80 pc) and an adaptive gas particle softening length with a minimum value of 1.5 pc. Star formation requires dense ($n \geqslant 300 \text{ cm}^{-3}$) molecular gas that is both self-shielding and Jeans unstable. Stellar feedback is implemented for Type Ia and Type II supernovae and stellar winds from OB and asymptotic giant branch stars. Radiative heating and pressure are included through photoionization and the photoelectric effect. This run does not include supermassive black hole feedback.

Note that although FIREbox has very high resolution for a cosmological volume simulation, it is 9–30 times lower resolution in baryonic mass than many FIRE zoom-in simulations (A. Wetzel et al. 2023). In this study, we focus on galaxies with stellar masses larger than $10^8 M_\odot$, which ensures more than ∼1600 star particles per galaxy.

We include only central galaxies and exclude any galaxies that are undergoing an interaction to limit environmental effects. We define an interaction as two galaxies with a stellar mass ratio less than 100 separated by less than $3 \times R_{80}$ (radius that encloses 80% of the stellar mass) of the more massive galaxy. Our final sample consists of 698 central galaxies broken up into 317, 273, and 108 objects in the stellar mass bins $10^{8-9}$, $10^{9-10}$, and $10^{10-11} M_\odot$, respectively.

### 2.2. Stellar Particle Mass Tensor

We measure the 3D shape of each galaxy using a mass tensor (S. Genel et al. 2015) as described in L. Y. Xia et al. (2025). Specifically, we calculate axes lengths $A$, $B$, and $C$ (from longest to shortest) using a weighted inertia tensor:

$$S_{ij} = \frac{\sum_n r_{\text{ell},n}^{-2} m_n x_{n,i} x_{n,j}}{\sum_n m_n}, \tag{1}$$

where $m_n$ represents the mass of each stellar particle, $x_{n,i}$ and $x_{n,j}$ are $i$- the and $j$-axis components of the particle positions ($x_1, x_2, x_3 = x, y, z$), and $r_{\text{ell},n}$ is the elliptical radius given by

$$r_{\text{ell}} = \sqrt{x^2 + \frac{y^2}{(B/A)^2} + \frac{z^2}{(C/A)^2}}. \tag{2}$$

The weight $r_{\text{ell},n}^{-2}$ mitigates the influence of particles located far from the galaxy's center. Diagonalizing the tensor yields eigenvalues proportional to the squares of the principal axes $A$, $B$, and $C$. We initially use stellar particles within a sphere ($A = B = C$) of radius $0.1R_{\rm vir}$, where $R_{\rm vir}$ is the virial radius as defined by G. L. Bryan & M. L. Norman (1998). The resultant tensor provides a new set of axis lengths. Including stars within the associated ellipsoid, we recalculate the tensor iteratively until the axis ratios $B/A$ and $C/A$ both converge within a value of $10^{-4}$.

It is subjective to define the "true" shape of a galaxy, and therefore worthwhile to explore multiple options. We defined the 3D shape using the tensor in three applications: All Stars, Young Stars, and Luminosity-weighted Stars in the $r$ band. For the All Stars and Young Stars measurements, we apply the mass-weighted tensor to the full stellar population and stellar population with ages $< 0.5$ Gyr, respectively. For the Luminosity-weighted Stars, we replace the particle mass ($m_n$) in Equation (1) with the particle luminosity ($l_n$) in $r$ band (based on the particle's mass, age, and metallicity; G. Chabrier 2003).

Each of these definitions give different insights into the galaxy. The All Stars shape provides an integrated measure of stellar shape. The Young Stars measure tells us about the geometric layout of star formation in an instantaneous manner, and allows us to ask questions like "do all stars form in disks?" (S. Yu et al. 2023). Finally, we explore the Luminosity-weighted shape because it may be more directly related to the observable shape.

### 2.3. Mock Images

We use the advanced Monte Carlo radiative transfer code SKIRT (P. Camps & M. Baes 2015, 2020) to generate mock images in the Sloan Digital Sky Survey (SDSS) $u$, $g$, and $r$ bands. The primary results presented in the following rely on the $r$-band images, which maximize our sample size (see Section 2.4).

SKIRT tracks the scattering, absorption, and extinction of photon packets emitted by radiation sources as the photons interact with a dust medium. SKIRT supports the implementation of both oligochromatic and panchromatic wavelength regimes, with the oligochromatic regime operating on a few distinct wavelengths and the panchromatic regime drawing from a user-specified wavelength grid that can support secondary thermal emission from dust and/or gas. Since we are interested in the optical range only, we use the extinction-only oligochromatic regime and use distinct wavelengths corresponding to the SDSS $u$, $g$, and $r$ bands.

For each gas particle, we assign the THEMIS dust mix (A. P. Jones et al. 2017) with silicate and hydrocarbon grain sizes of eight bins. SKIRT uses the parameter $f_{\rm mass}$, known as the dust-to-metal ratio, to calculate the dust mass from the gas mass:

$$M_{\rm dust} = f_{\rm mass} Z M_{\rm gas}, \tag{3}$$

where $Z$ is the gas metallicity. For our analysis, we set $f_{\rm mass}$ to 0.3 for $10^8$–$10^{10}$ $M_\odot$ galaxies (V. Rodriguez-Gomez et al. 2019). At higher stellar masses, $10^{10}$–$10^{11}$ $M_\odot$, galaxies in FIREbox have gas-phase metallicities that are a factor of $\sim$3 too high compared to observations (R. Feldmann et al. 2023). For this reason, we use $f_{\rm mass} = 0.1$ in this mass range as our fiducial choice. This has the same effect as lowering the fraction of dust, which corrects for the excess metallicities in our galaxies at this mass. Note that at lower masses the gas-phase metallicities in FIREbox are in excellent agreement with observations (R. Feldmann et al. 2023).

Finally, we assign the stellar population model of G. Bruzual & S. Charlot (2003) with a Chabrier initial mass function to each star particle. Each particle in the simulation is then smoothed using a scaled Gaussian smoothing kernel with a smoothing length of 1.4 times the particle's gravitational softening length. For galaxies with $M_\star < 10^{10}$ $M_\odot$, we set the field of view (FOV) to $4.5R_{80}$. For galaxies with $M_\star > 10^{10}$ $M_\odot$ the FOV is $0.4R_{\rm vir}$, which ensures a capture of the full stellar structure.

To resolve high-density regions, we set the minimum grid level in our octree grid to four. Additionally, we fix a spatial resolution of $400 \times 400$ pixels for images with an FOV greater than 50 kpc and $300 \times 300$ pixels for images with an FOV less than 50 kpc. As greater resolutions require a greater number of photon packets to ensure all areas are illuminated, we set a fixed photon resolution of 50 photon packets per pixel, which corresponds to $8 \times 10^6$ and $4.5 \times 10^6$ photon packets for FOVs greater than and less than 50 kpc, respectively. Another option is to set a pixel scale of 40 pc, twice the separation of gas particles eligible for star formation (R. Feldmann et al. 2023). However, this is computationally expensive to implement for 698 galaxies, and we find that our pixel scale does not significantly impact the morphological measurements within the bounds of this paper.

We create projected images along random and principal-axis orientations (PAOs). Specifically, we create three randomly oriented images of each galaxy by projecting along each of the three coordinate axes of the simulation box. This yields a total sample of 2094 randomly oriented galaxy images. The PAO images are created by projecting along the principle axes of the galaxy determined in Section 2.2 using the All Stars mass tensor. Projecting along the smallest principal axis, $C$, creates the equivalent of a face-on image. Alternatively, projecting along the intermediate $B$ axis is equivalent to an edge-on view.

We also create a set of projected stellar mass–surface density "images" along the same orientations as the mock observations. Compared with the full mock images, these mass–density projections provide a comparison set that helps us to understand the effects of dust and mass-to-light ratio variations on the inferred galaxy properties. Throughout this paper, we define our total stellar mass as the stellar mass contained within the galaxy's FOV, which is approximately equal to the total bound stellar mass for all galaxies.

Figure 1 provides three example $ugr$ composite PAO images of three representative galaxies, with stellar mass increasing from top to bottom. Each row shows a single galaxy viewed along its short, long, and intermediate axis, from left to right.

### 2.4. Sérsic Fitting

Our primary method for characterizing the 2D shapes of our simulated galaxies is by using AstroPhot (C. J. Stone et al. 2023) to fit a Sérsic surface brightness profile (J. L. Sérsic 1963) to our mock images:

$$I(r(x, y)) = I_e \exp\left\{-b_n\left[\left(\frac{r}{R_e}\right)^{1/n} - 1\right]\right\}, \tag{4}$$

where $r(x, y)$ is a rotated elliptical radius defined by an orientation angle and a minor-to-major axis ratio $q \equiv b/a$. $I_e$ is the surface brightness at $R_e$. The Sérsic index is $n$ and $b_n$ is a constant dependent on $n$ defined such that $R_e$ contains half of the integrated light. The $q$ values listed in each panel of Figure 1 provide example axis ratios derived this way. Note that when galaxies are projected along the axis perpendicular to their longest and shortest 3D axes (right column of Figure 1) the inferred 2D axis ratios should be the smallest for that galaxy, although there can be a small variance especially in more spheroidal galaxies.

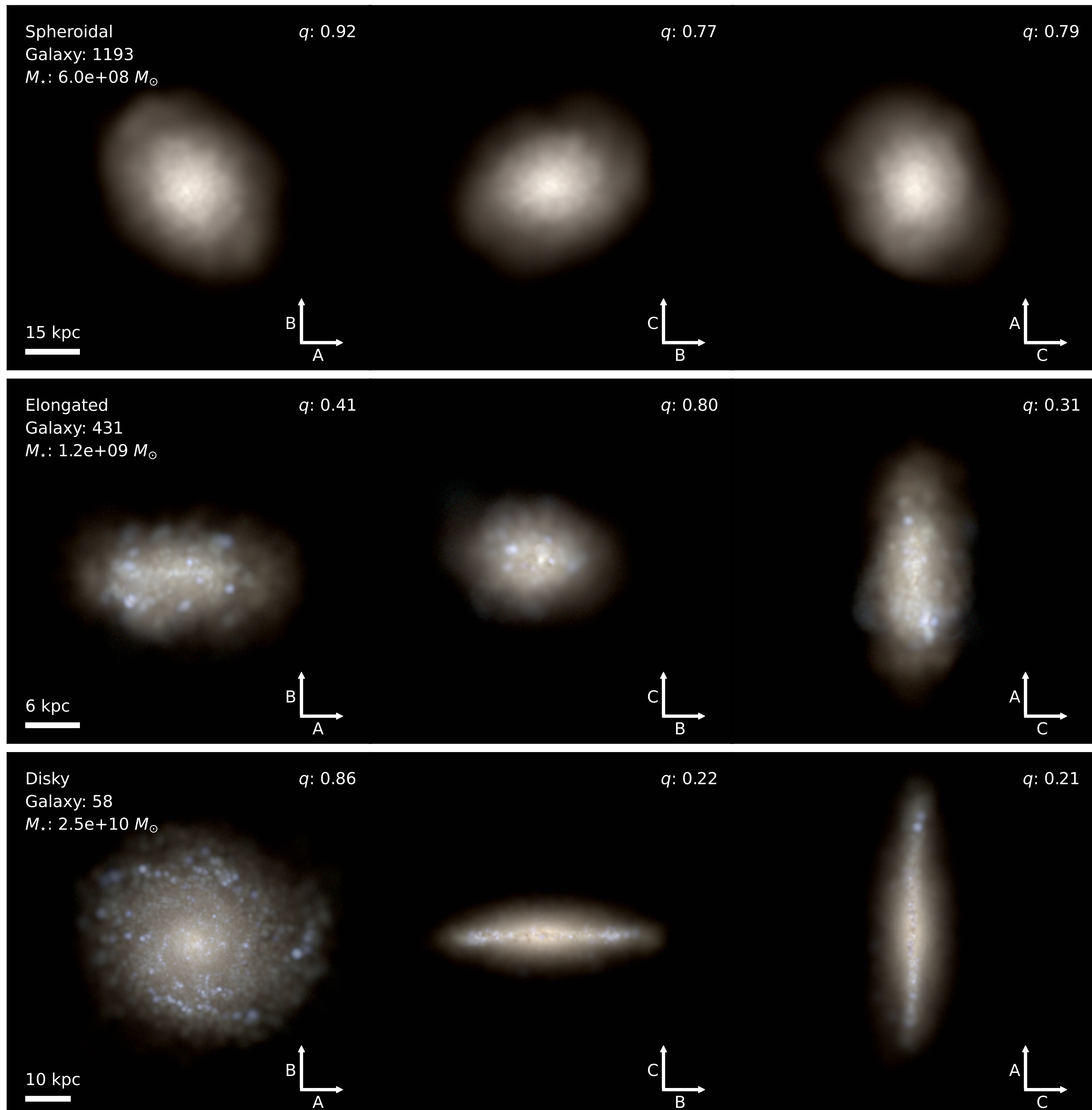

**Figure 1.** Example set of *ugr* composite mock images with `SKIRT` for three simulated galaxies with representative 3D shapes: spheroidal (top row), elongated (middle), and disky (bottom). The three images in each row are oriented along the short ($C$), long ($A$), and middle ($B$) axis derived from the mass-weighted All Stars tensor. The two remaining axes that define the plane of view of each image are indicated in the bottom right. The measured 2D axis ratio, $q$, from an $r$-band Sérsic fit in each orientation is listed in the upper right of each image. The spheroid in the top row is from the $10^8$–$10^9$ $M_\odot$ stellar mass bin and has 3D All Stars axis ratios ($B/A$, $C/B$, $C/A$) = (0.97, 0.91, 0.89). The elongated example is from the $10^9$–$10^{10}$ $M_\odot$ mass bin and has 3D axis ratios (0.59, 0.81, 0.48). The disky example along the bottom row is from the $10^{10}$–$10^{11}$ $M_\odot$ mass bin and has 3D axis ratios (0.94, 0.34, 0.32).

In about 1% of the projections in the $r$ band, isolated bursts of star formation affect our ability to characterize the light profile with a physically meaningful 2D Sérsic profile. For the $g$ and $u$ bands this occurred in 3% and 9% of the images, respectively. Most of these cases involve single star-forming regions that dominate the fit such that $R_e < 1$ kpc and $q \simeq 1$. We have excluded these fits from our analysis, leaving us with approximately 2000 good axis ratio measurements in the $r$ band. More specifically, we have 927, 819, and 324 measured axis ratios in the stellar mass bins $10^{8-9}$, $10^{9-10}$, and $10^{10-11}$ $M_\odot$, respectively.

### 2.5. GAMA Sample

We compare our mock images to results from the GAMA survey (S. P. Driver et al. 2009, 2011) Data Release 3 (DR3; I. K. Baldry et al. 2018). We use the stellar masses from the StellarMassesv20 catalog (E. N. Taylor et al. 2011). We apply the recommended flux scale correction, which accounts for the conversion of the flux measured with the $r$-defined AUTO photometry to the flux inferred with the Sérsic profile fit, which is truncated at $10R_e$ (E. N. Taylor et al. 2011). Following M. Haslbauer et al. (2022), we make a quality cut requiring $0.5 < \text{flux scale} < 2$ and the posterior predictive $p$-value for the spectral energy distribution (SED) fit to be greater than zero. We are using the Sérsic $r$-band measurements from the SérsicCatSDSSv09 catalog (L. S. Kelvin et al. 2012), which applied a single 2D Sérsic profile using GALFIT 3 (C. Y. Peng et al. 2010). We use the $r$-band ellipticity $(1 - b/a)$ measured from the Sérsic fit to determine the observed axis ratio ($q = b/a$) for the GAMA sample. We use galaxies with a corrected stellar masses in the range $10^8$–$10^{11}$ $M_\odot$ within a redshift range of $0.005 < z < 0.04$. This provides a sample size comparable to our simulated mock observation sample, with 1216, 566, and 304 galaxies in the stellar mass bins $10^{8-9}$, $10^{9-10}$, and $10^{10-11}$ $M_\odot$, respectively.

## 3. Results

### 3.1. Intrinsic Three-dimensional vs. Two-dimensional Projected Morphologies

In this section, we compare the intrinsic 3D morphologies of our simulated FIREbox galaxies to inferred shape measurements from idealized 2D projections. Figure 2 presents our results in the space of short-to-long ($C/A$) and intermediate-to-long ($B/A$) axis ratios. We follow A. van der Wel et al. (2014) and divide the parameter space into characteristic morphological regions: spheroidal, disky, and elongated. These regions are labeled in the left column panels. Qualitatively, spheroidal galaxies have $A \approx B \approx C$, elongated galaxies have $A > B \approx C$, and disky galaxies have $A \approx B > C$. Examples of galaxies that fall into each of these three categories are shown from top to bottom in Figure 1.

The columns in Figure 2 subdivide our sample into three stellar mass bins: $10^8$–$10^9 M_\odot$, $10^9$–$10^{10} M_\odot$, and $10^{10}$–$10^{11} M_\odot$, from left to right. The upper three rows of Figure 2 show 3D axis ratios derived using the tensor analysis described in Section 2.2. The top row presents shapes derived using the mass-weighted tensor of all star particles (All Stars), and the second row shows results using the mass-weighted tensor of stars younger than 0.5 Gyr (Young Stars). The third row shows results for the shape of all star particles weighted by their luminosity in the $r$ band (Luminosity-weighted Stars).

The bottom row presents the same galaxies, now in a quasi-observable way. Each galaxy is mock observed twice to determine an idealized "observable" $B/A$ and $C/A$. Using the axes determined by the All Stars mass tensor, first the galaxy is projected along its short axis to be viewed "face on," as in the left panels of Figure 1. We then perform a Sérsic $r$-band fit to determine its observed axis ratio, $q_{\text{face-on}}$, and associate this with the $B/A$ axis ratio. Second, we project the same galaxy along its intermediate axis to view it "edge on," as in the right panels of Figure 1. We associate the resultant Sérsic fit axis ratio, $q_{\text{edge-on}}$, with $C/A$.

In all four rows, we see a gradual shift toward smaller $C/A$ ratios and larger $B/A$ ratios as we progress to higher masses. The Young Stars show the most morphological variation as a function of galaxy stellar mass. At high masses, young stars reside in disky configurations, while in the intermediate- and lower-mass bins they shift toward elongated and/or spheroidal configurations. The $r$-band Luminosity-weighted Stars and PAO mock shapes are qualitatively intermediate between the All Stars and Young Stars. When compared to the Luminosity-weighted Stars shapes, the mock images have objects that are more spherical in the highest-mass bin.

Figure 3 and Table A1 provide quantitative summaries of the shape data presented in Figure 2. The columns in Figure 3, from left to right, present shape statistics for All Stars, Young Stars, Luminosity-weighted Stars, and the PAO $r$-band mock images, respectively. Each panel in the top row summarizes the fraction of galaxy shapes classified as spheroidal (green), elongated (blue), and disky (red) as a function of stellar mass. Panels in the second row of Figure 3 show the distribution of galaxy triaxiality ($T$) as a function of stellar mass, where $T$ is defined as in M. Franx et al. (1991):

$$T = \frac{A^2 - B^2}{A^2 - C^2}. \tag{5}$$

Panels in the third row of Figure 3 show the distribution of ellipticity parameters ($E = 1 - C/A$) as a function of stellar mass. Triaxiality and ellipticity are the two main parameters used to describe the ellipsoid derived from an axis ratio distribution, therefore we can directly compare our 3D shapes to those derived from observations. In both the second and third rows, solid colored lines represent the median values in the relevant mass bins, and the shaded colored boxes denote the inner quartiles. Additionally, the whiskers denote the outer quartiles and circles are outliers.

The top left panel of Figure 3 shows that the All Stars shapes are overwhelmingly spheroidal at low and intermediate masses and that roughly half are disky at high masses (Milky Way scale). As seen in the middle and bottom left panels, the median triaxiality decreases and the median ellipticity increases systematically with increasing stellar mass. This is interesting in light of the top panel: even though most of the intermediate-mass galaxies are classified as spheroidal, those spheroids are more oblate than the spheroids at lower masses. In the highest-mass bin, we see significantly lower $T$ values and higher ellipticities; this is consistent with the emergence of a disky population at these high masses.

The Young Stars morphological variation as a function of stellar mass is significantly more pronounced when compared to the All Stars (Figure 3, top of second column). In the lowest-mass bin, about 80% of galaxies have their Young Stars arranged in spheroidal shapes. At intermediate masses, roughly half of the galaxies have Young Stars arranged in elongated shapes. In about 90% of the high-mass galaxies, Young Stars are arranged in disky configurations. This is fairly remarkable, as while star formation largely occurs in disks at high mass, this is not the case at lower- and intermediate-mass galaxies. Instead, star formation appears to proceed in more elongated configurations at intermediate mass and more isotropically at low mass. In the second row of the second column we see that the triaxiality parameter is the same for low- and intermediate-mass galaxies, indicating the parameter was not sensitive to the shift in morphology, although in the third row we see that the ellipticity of Young Stars increases systematically with increasing galaxy mass, as would be expected for the shift to more elongated and disky configurations.

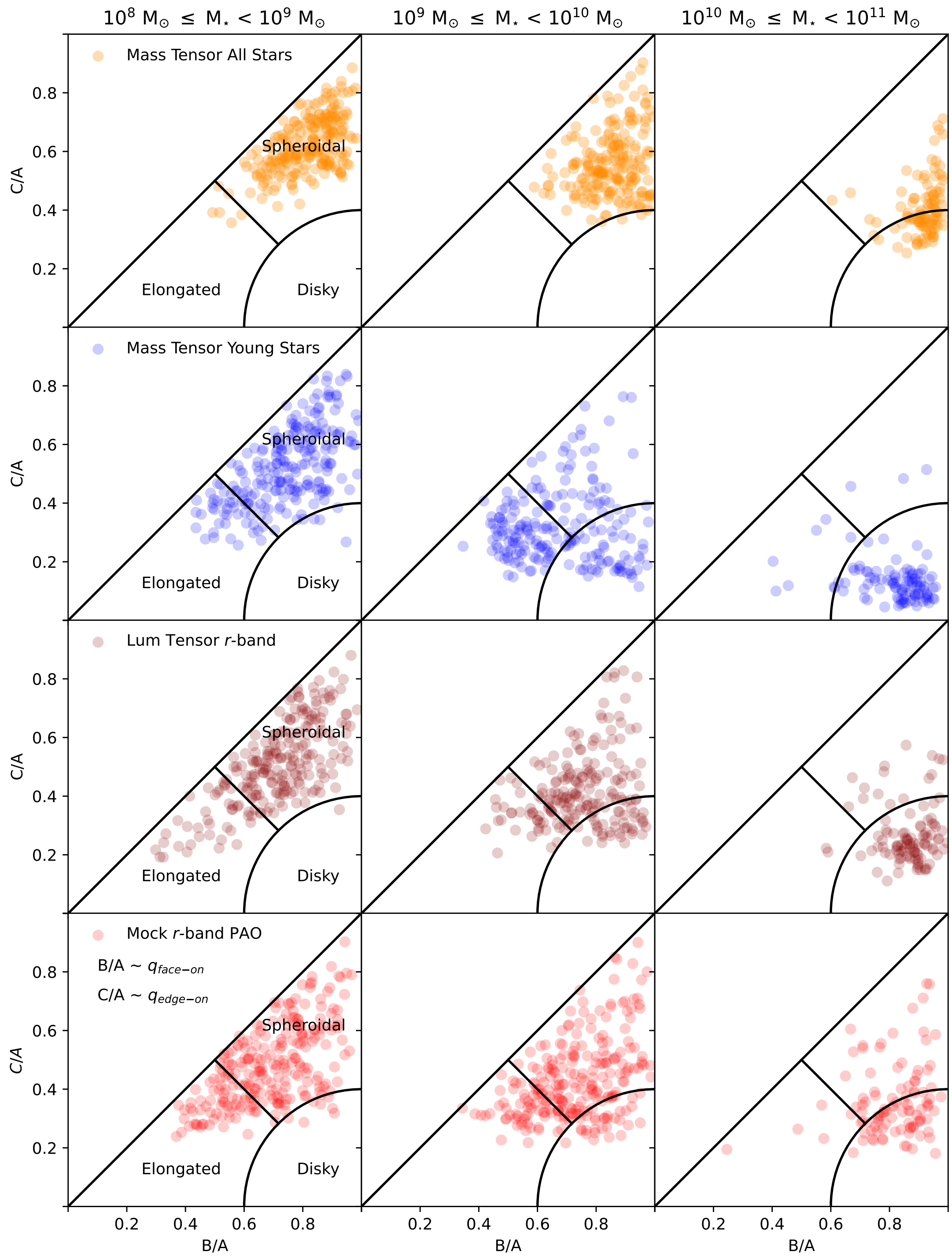

**Figure 2.** Predicted galaxy 3D axis ratios, assuming each galaxy is well approximated by a triaxial ellipsoid with axes $A > B > C$. Categorical shapes are indicated by the defined regions in each panel. The three columns display the results for galaxy stellar mass bins $10^8$–$10^9$ $M_\odot$, $10^9$–$10^{10}$ $M_\odot$, and $10^{10}$–$10^{11}$ $M_\odot$, from left to right. The first three rows show the axis ratios, $B/A$ vs. $C/A$, calculated using the weighted inertia tensors of All Stars, Young Stars (<0.5 Gyr), and *r*-band Luminosity-weighted Stars, respectively. In all cases, we see more spheroidal galaxies at lower masses and more disky galaxies at higher masses, though the differences are more pronounced in Young Stars. The bottom row shows a quasi-observable representation of galaxy shapes based on "observed" axis ratios along two orientations for each galaxy. Specifically, for each galaxy, we assign a $B/A$ value using the 2D axis ratio, $q_{\text{face-on}} = b/a$, derived from a Sérsic fit to a face-on *r*-band mock image of the galaxy (oriented along the short axis as defined by its All Stars inertia tensor, as in the left most panels of Figure 1). We assign the same galaxy a $C/A$ value using the 2D axis ratio, $q_{\text{edge-on}} = b/a$, derived from a Sérsic fit to the edge-on *r*-band mock image (as in the right most panels of Figure 1).

The shape distributions for the Luminosity-weighted Stars tensor and PAO observations are fairly similar (first row of the third and fourth columns of Figure 3). The lowest-mass bin for both metrics has ∼75% spheroidal configurations and ∼20% elongated. At intermediate masses, both measures have slightly smaller spheroidal fractions (∼60%) and nonnegligible disky populations (∼15%). In the highest-mass bin, the Luminosity-weighted Stars shapes are 84% disky, while the PAO shapes are only 56% disky. As a function of stellar mass, both the Luminosity-weighted Stars shapes and PAO shapes show decreasing triaxiality and increasing ellipticity as a function of increasing stellar mass.

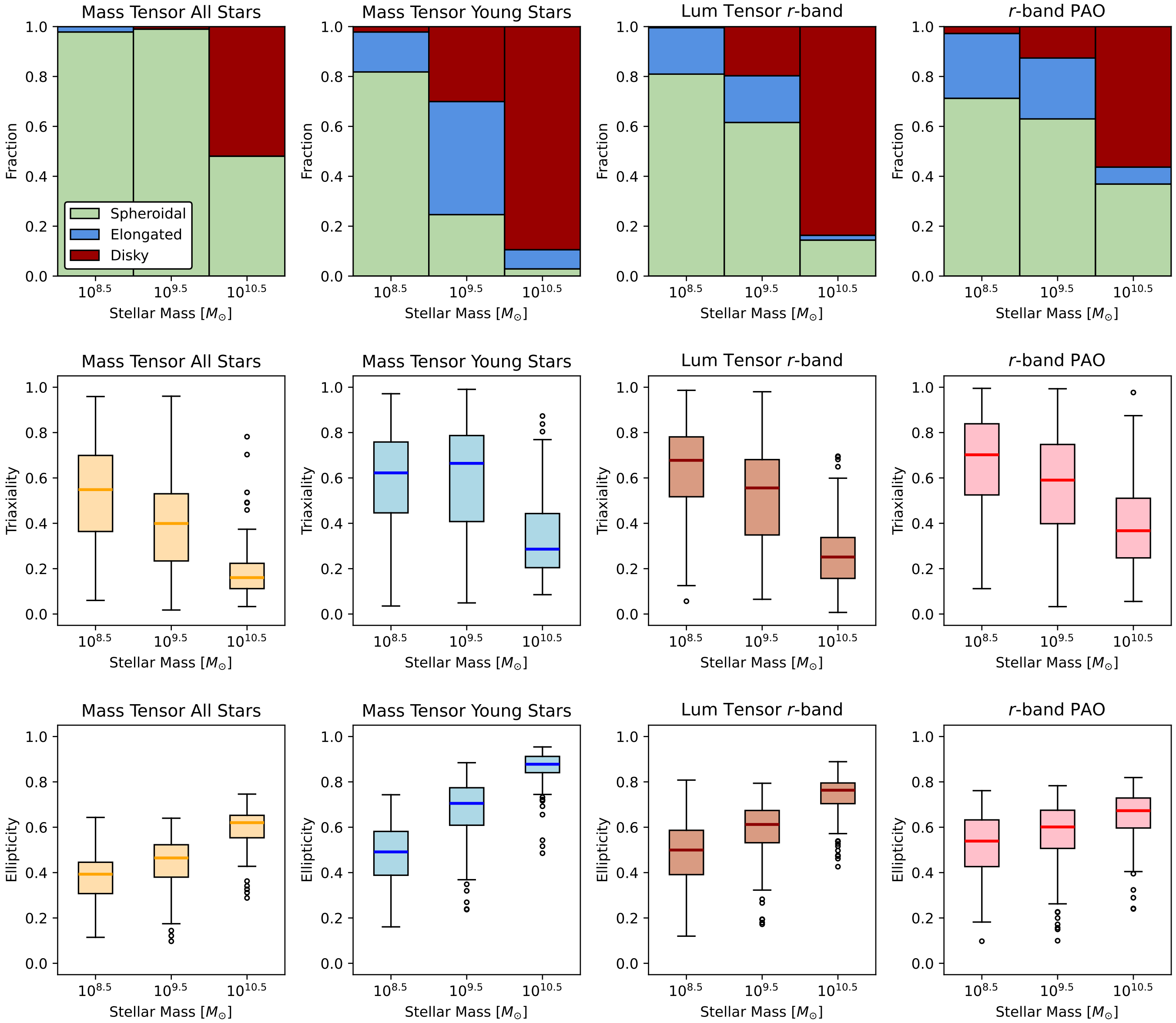

**Figure 3.** Row 1: the fraction of spheroidal (green), elongated (blue), and disky (red) galaxies derived from Figure 2 with each panel showing the results for our four methods: All Stars mass tensor, Young Stars mass tensor, *r*-band Luminosity-weighted Stars tensor, and *r*-band PAO images. Each column in the panel shows the composition of morphologies for the given stellar mass bins: $10^8$–$10^9$ $M_\odot$, $10^9$–$10^{10}$ $M_\odot$, and $10^{10}$–$10^{11}$ $M_\odot$. Row 2: the triaxiality of the galaxies defined by Equation (5) with each panel showing the results for All Stars mass tensor (orange), Young Stars mass tensor (blue), *r*-band Luminosity-weighted Stars tensor (dark red), and *r*-band PAO images (light red), with each plot within the panel representing the given stellar mass bin. The solid colored line is the median triaxiality, with the box representing the inner two quartiles, and the bars represent the outer two quartiles, with dots representing outliers. Row 3: same as row 2, except showing ellipticity.

Our exploration of the quasi-observable PAO shape distributions is motivated by an effort to understand which of our "true" 3D shape distributions would be most closely matched to an idealized observational sample of only edge-on and face-on galaxies. As discussed in the paragraph above and also in Appendix A with Figure A1, the observable shape distribution most resembles the Luminosity-weighted Stars 3D shape distribution and differs the most from All Stars stellar-mass-weighted 3D shape distribution. While this is perhaps not surprising, we note that the 3D shapes derived from a 2D axis ratio distribution in observations are often interpreted as the shape of the total stellar population.

### *3.2. Predicted Two-dimensional Shapes vs. Observations*

In this section, we apply 2D Sérsic fits to randomly oriented mock *r*-band images and compare the resulting shape measurements to those from the GAMA DR3 observational sample.

Figure 4 focuses on galaxies with stellar masses similar to the Milky Way: $10^{10} M_\odot < M_\star < 10^{11} M_\odot$. The black line shows the observed distribution, and the solid red histogram shows our predicted distribution. For reference, the dashed red histogram shows the shape distribution derived using the stellar mass–surface density images of the same galaxies. Note that mass-derived 2D shapes are rounder on the sky than the more realistic mock images, highlighting the importance of the mock image comparison.

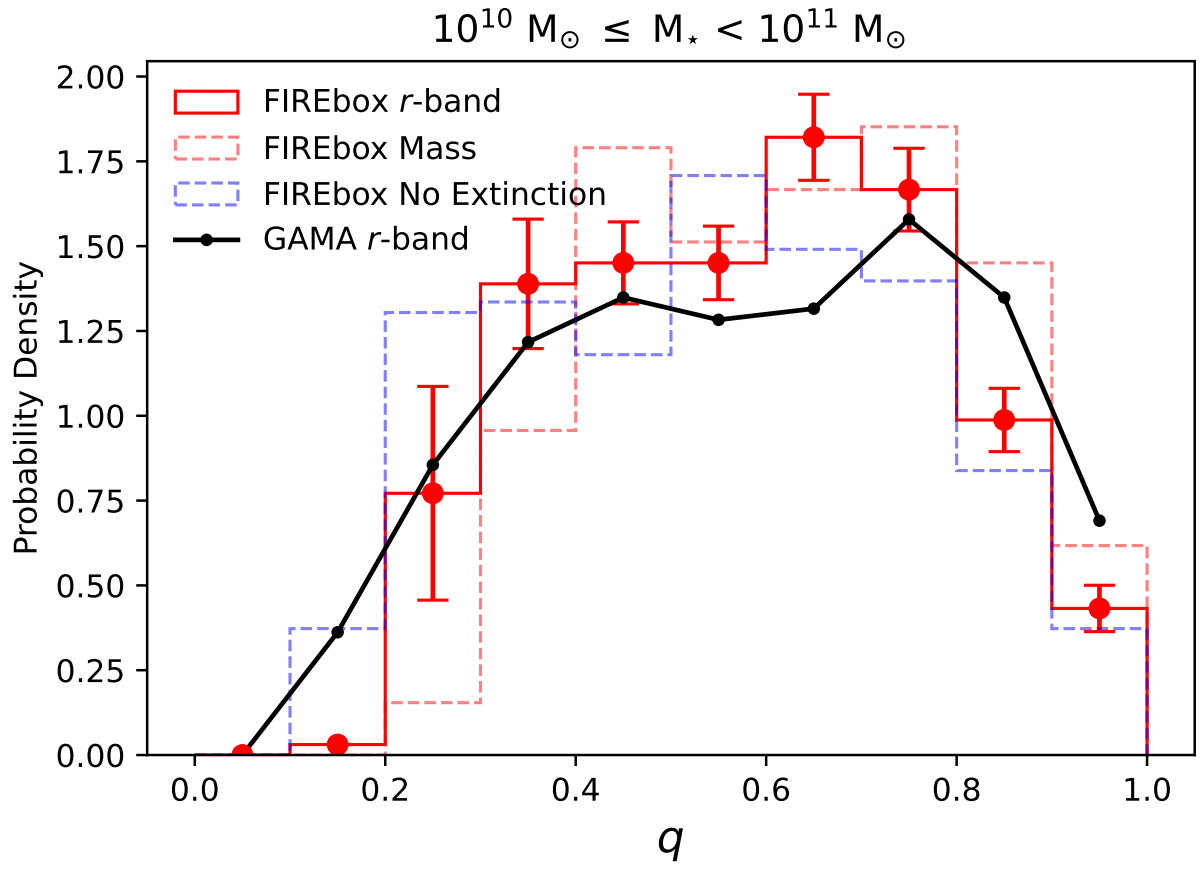

**Figure 4.** Observed and predicted 2D axis ratio distributions for galaxies of $10^{10}\,M_\odot < M_\star < 10^{11} M_\odot$. The solid black line is the $q$ distribution measured using Sérsic $r$-band fits to real galaxies from GAMA DR3. The solid red histogram show the $q$ distribution for Sérsic fits to randomly oriented $r$-band mock images of FIREbox galaxies. The error bars are Poisson counting errors for each bin. The dashed red histogram shows the $q$ distribution for FIREbox galaxies derived from (nonobservable) stellar mass–surface density Sérsic fits. The dashed blue histogram shows the $q$ distribution for FIREbox galaxies without dust altogether. Note that while the mass-projected shapes are not a good match to the observed distribution, the more realistic mock images produce a distribution of axis ratios that are moderately similar to that observed.

The predicted FIREbox distribution matches the GAMA sample relatively well between $0.2 < q < 0.8$. In particular, the simulation is capable of producing a reasonable fraction of $q \sim 0.5$ galaxies compared to observations, but there is still a lack of thin galaxies ($q \sim 0.15$). A Kolmogorov–Smirnov (K-S) statistical comparison between the two distributions, as shown in Table 1, yields a $p$-value of 0.45. This result supports the possibility that the two distributions are sampled from the same underlying distribution.

However, this result holds tension as the shape distribution is heavily dependent on dust. As discussed in Section 2.3, the dust-to-metal ratio in SKIRT, $f_{\rm mass}$, is the value multiplied by the gas metallicity and gas mass to obtain the dust mass, where $M_{\rm dust} = f_{\rm mass} Z M_{\rm gas}$. Our results so far are based on the assumption that $f_{\rm mass} = 0.1$ for high-mass FIREbox galaxies. As explored in Appendix D, galaxy shapes are highly sensitive to the dust-to-metal ratio. Raising $f_{\rm mass}$ to 0.3 tends to thicken edge-on galaxies. With this choice, we would obtain a significantly lower $p$-value of 0.18 between the FIREbox and GAMA distributions.

Figure 5 compares the predicted and observed axis ratio distributions for galaxies at lower stellar masses: $10^8 M_\odot < M_\star < 10^9 M_\odot$ (left) and $10^9 M_\odot < M_\star < 10^{10} M_\odot$ (right). As in Figure 4, the solid red histograms show the results for our randomly oriented $r$-band Sérsic $q$ measurements and the solid black lines show the observed GAMA $r$-band axis ratios for the same mass bins. In both mass bins, the predicted distribution has far too few galaxies with $q < 0.3$. As presented in Table 1, the K-S tests yield $p < 10^{-5}$ in both cases, indicating statistically significant differences between the distributions.

To further emphasize the problem revealed in Figure 5, the dashed red histograms show the $q$ distributions for FIREbox galaxies viewed edge on, such that $q \sim C/A$. We see that there are no galaxies in the simulation that are as thin as the thinnest galaxies observed. In principle, the problem could be that the simulation produces too few disky systems with $C/A \lesssim 0.2$ or too few elongated systems with $C/A \lesssim 0.2$. In our simulation, the (higher-mass) galaxies that have edge-on axis ratios $< 0.2$ are always disks (bottom row, Figure 2). This suggests that the mismatch we see here relates to an underproduction of disk galaxies at low masses in our simulation.

**Table 1**
Kolmogorov–Smirnov Statistic for the Difference between the $r$-band $q$ Distribution in FIREbox and GAMA (Shown in Figures 4 and 5)

| Mass ($M_\odot$) | K-S Statistic | $p$-value |
|---|---|---|
| $10^8$–$10^9$ | 0.19 | $<10^{-5}$ |
| $10^9$–$10^{10}$ | 0.17 | $<10^{-5}$ |
| $10^{10}$–$10^{11}$ | 0.067 | 0.45 |

**Note.** Mass indicates the mass bin, the K-S statistic quantifies the difference between the two distributions, and the $p$-value quantifies the likelihood of the two distributions being sampled from the same underlying distribution.

## 4. Discussion and Conclusions

We have used the FIREbox simulation to study the 3D shapes of galaxies with stellar masses of $10^8$–$10^{11} M_\odot$ and to connect those shapes to observable axis ratios using mock images (see Figure 1).

Under the assumption that our simulated galaxies are well approximated by triaxial ellipsoids, we measured the axis ratios of subpopulations in each galaxy: All Stars, Young Stars, and Luminosity-weighted Stars (see Figures 2 and 3). We find that Young Stars (<0.5 Gyr), in particular, are arranged in systematically different configurations as a function of galaxy stellar mass. In ∼90% of high-mass galaxies ($M_\star \sim 10^{10.5} M_\odot$), the Young Stars reside in disky (oblate) configurations. In contrast, the plurality (∼50%) of intermediate-mass galaxies ($M_\star \sim 10^{9.5} M_\odot$) have Young Stars in elongated (prolate) shapes, while the large majority (∼80%) of low-mass galaxies ($M_\star \sim 10^{8.5} M_\odot$) have Young Stars in spheroidal configurations. The prevalence of elongated shapes among star-forming material at intermediate masses in our simulation is interesting in light of observational evidence that prolate systems are common among star-forming galaxies in the early Universe (V. Pandya et al. 2024).

The full stellar-mass-weighted shapes of our galaxies are generally more spherical than the Young Stars shapes. On average, the All Stars shapes have decreasing triaxiality and increasing ellipticity as stellar mass increases (left column, Figure 3). Our $r$-band Luminosity-weighted Stars shapes are intermediate between the Young Stars and All Star shapes.

As a step toward understanding how the observable 2D shapes of galaxies are related to intrinsic 3D shapes, we used idealized "edge-on" and "face-on" views of each galaxy to compare the associated 2D axis ratio pairs with the underlying 3D axis ratios ($q_{\rm face\text{-}on}$ to $B/A$ and $q_{\rm edge\text{-}on}$ to $C/A$). We find that the 3D shapes derived from Luminosity-weighted Stars most closely match the inferred mock shapes, as expected, given that both are based on the $r$-band luminosity. However, the All Stars mass-weighted shapes are typically much more spheroidal than would be implied by the mock images (see Figure A1). Often, the 3D shapes derived from 2D axis ratio distributions from observations are assumed to represent the stellar-mass-weighted 3D shape. Our result suggests that when we observe and measure galaxy shape distributions on the sky, we are in a better position to constrain the luminosity-weighted 3D shapes of galaxies than the 3D stellar-mass-weighted shapes. If comparing 3D shapes between observations and simulations, the simulated 3D should be luminosity weighted.

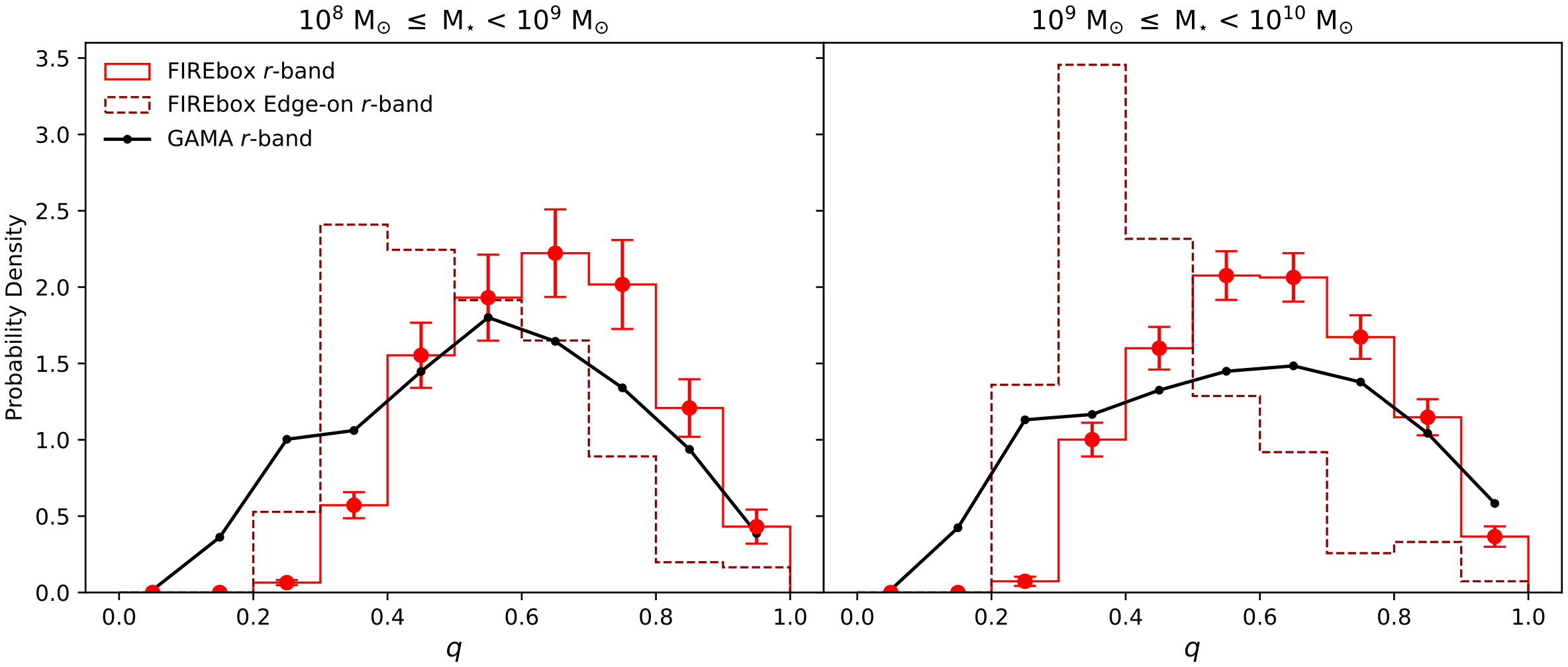

**Figure 5.** 2D axis ratio distributions for galaxies $10^8\ M_\odot < M_\star < 10^9 M_\odot$ (left) and $10^9\ M_\odot < M_\star < 10^{10}\ M_\odot$ (right). The solid red histogram is the $q$ distribution for the randomly oriented FIREbox galaxies. The error bars show the Poisson errors for reference. The dark red dashed histogram shows the $q$ distribution for the same galaxies oriented edge on. The solid black line is the $q$ distribution from GAMA DR3 for the same mass bins. For both mass bins, there are fewer predicted galaxies with $q < 0.3$ than seen in observations.

We then compared the predicted on-sky shape distributions of our galaxies (now randomly oriented) to those observed using GAMA DR3 results. We found that for massive, Milky Way–size galaxies, the simulated galaxy shape distribution is a reasonably good match to the observations (Figure 4) but heavily depends on the dust-to-metal ratio. In particular, in the stellar mass range of $10^{10-11} M_\odot$, FIREbox produces a sufficient population of galaxies that are as thin as $q \sim 0.25$. With no dust, this lower bound goes down to $q \sim 0.15$, the thinnest disks observed in the sky. Importantly, this would not be the case had we used the mass-projected shapes instead of realistic mock images (see dashed line in Figure 4), or if we used a higher dust-to-metal ratio, stressing the importance of an empirically derived dust population. Using our current constant dust-to-metal ratio of 0.1, this result is consistent with the findings of D. Xu et al. (2024), who used the Illustris TNG50 simulations in a similar analysis.

Although our predicted thin galaxy fraction at the Milky Way scale is in reasonable agreement with observations, the predicted shape distributions at lower masses are significantly different than observed. As seen in Figure 5, our axis ratio distributions are well below the GAMA distributions for $q < 0.3$. We interpret this as an underprediction in the disk galaxy fraction at stellar masses below $10^{10}\ M_\odot$. This interpretation is consistent with the reports of A. van der Wel et al. (2014) who used $g$-band SDSS photometry and an associated triaxial ellipsoid model to conclude that over 80% of $10^9$–$10^{11}\ M_\odot$ star-forming galaxies are disks at $z = 0$. This is far above our predicted disk fraction in this mass range.

The mismatch presented in Figure 5 is consistent with the findings of K. El-Badry et al. (2018a, 2018b), who studied the emergence of disky morphology in a small number of high-resolution FIRE-2 simulations. Focusing on H I kinematics in low-mass galaxies, they found that some H I gaseous disks appear in observed low-mass galaxies at an order-of-magnitude lower stellar mass scale than any of the FIRE-2 simulations studied in that paper. Taken together with our findings, this mismatch could indicate that "churning" from stellar feedback in FIRE-2 may be too vigorous, puffing up the galaxies. This explanation is supported by the apparent overprediction of star formation rate burstiness in FIRE at this mass scale (M. Sparre et al. 2017; N. Emami et al. 2019).

Unfortunately, this explanation may be too simplistic as there is no theoretical consensus on the key physical drivers of galaxy disk formation. For example, a correlation between the virialization of the inner circumgalactic medium (CGM) and the onset of thin-disk formation has been identified within FIRE simulations (J. Stern et al. 2021; S. Yu et al. 2021; A. B. Gurvich et al. 2022; Z. Hafen et al. 2022). This may suggest that models with CGM virialization at a lower mass scale (by having a lower density, for example) could produce more low-mass disks. Similarly, higher central mass concentrations in simulated galaxies also correlate with disk formation (P. F. Hopkins et al. 2023), which help explain why some dissipative dark matter models tend to have more rotationally supported galaxies (X. Shen et al. 2021).

Alternatively, our underprediction in the disk galaxy fraction at lower masses could be related to resolution. In the higher-mass bin ($M_\star \sim 10^{10.5}\ M_\odot$), where the thin galaxy fraction looks good, we are able to resolve short axis lengths as small as $C \sim 0.1A \sim 0.1R_e \sim 400$ pc. For a typical galaxy with a lower mass, $M_\star \sim 10^9 M_\odot$, we would need to resolve its structure at ∼300 pc to produce a galaxy as thin as observed. Although the physical scale is similar, a galaxy of this stellar mass has ∼30 times fewer particles, which could produce a numerical effect. As a point of comparison, we have looked at the eight zoom-in simulation galaxies presented in C. Klein et al. (2024) with stellar masses in the relevant range of $10^8$–$10^{10}\ M_\odot$. Like FIREbox, these simulations use FIRE-2 physics, but they have 30 times higher mass resolution. 88% of the measured randomly oriented mocks images have axis ratios with $q > 0.4$, with the smallest measured value being 0.23. This is consistent with the 90% of mock images with $q > 0.4$ found in FIREbox, suggesting that resolution may not be the cause of the tension. However, given the limited number of zoom simulations available in this mass range, it is difficult to draw strong conclusions. This motivates the need for a larger set of high-resolution simulations in this mass range.

It is interesting to compare what we have found here using FIREbox galaxies and the results of D. Xu et al. (2024), who used mock observations of TNG50 galaxies to compare their shapes directly to Hyper Suprime-Cam Subaru Strategic Program Survey images in the *g* band, *i* band, and *y* band. As mentioned in Section 1, they concluded that there is no tension related to the shape distributions of Milky Way–mass galaxies, as previously suggested. They also find good agreement in the low-mass regime ($<10^{9.5}\,M_{\odot}$) in the *g* band, but as they progress toward longer wavelengths, they see distributions skewed toward larger axis ratios compared with the data, especially in the *y* band. Like us, they concluded that lower-mass galaxies in TNG50 were “intrinsically rounder/thicker than their observed counterpart,” though our results were focused on *r*-band comparisons. They suggest that spurious heating could play a role in their simulations. The better agreement they find in the optical bands is interesting, and, given the strong dependence we find on the shape distributions and the handling of dust, one potentially significant difference in their approach (which relied on C. Bottrell et al. 2024) is that their dust attenuation and scattering model assumes a metallicity-dependent dust-to-metal ratio. Their assumptions works out to be slightly lower dust fractions than ours at higher stellar masses.

The results presented here provide a window into how the observable 2D shapes of galaxies on the sky relate to the underlying 3D shapes of the stellar populations weighted in various ways. The FIREbox simulation appears to be successful at predicting the 2D shapes of galaxies at the Milky Way–mass scale, suggesting that the Luminosity-weighted Stars tensor stellar distributions are a good representation of the real Universe. However, the relative lack of thin, disky galaxies in this simulation at lower stellar masses highlights the need to test revised implementations of physics. One promising case is the upcoming `Azahar` simulations (T. Dome et al. 2025; S. Martin-Alvarez et al. 2026), which include population-level statistics and internal structure. These, along with studying the effects of dust on shapes, will help improve our understanding of the physical processes driving galaxy morphology. Furthermore, higher-resolution simulations, coupled with observational surveys from Euclid, the Vera C. Rubin Observatory's Legacy Survey of Space and Time, and other programs will greatly expand the sample of galaxies in this mass range at low redshift and provide an important way forward.

## Acknowledgments

We acknowledge past and continuing efforts to uplift and classify galaxies as disky, elongated, or isotropic (DEI). Such efforts related to DEI have been instrumental to the success of this paper and remain important for improving our understanding of the Universe. C.K. is supported by a National Science Foundation (NSF) Graduate Research Fellowship Program under grant DGE-1839285. J.S.B., L.Y.X., and J.D. W. are supported by NSF grants AST-1910965 and AST-2408247, and NASA grant 80NSSC22K0827. F.J.M. is funded by the NSF Math and Physical Sciences (MPS) Award AST-2316748. C.A.F.G. is supported by NSF through grants AST-2108230 and AST-2307327; by NASA through grants 21-ATP21-0036 and 23-ATP23-0008; and by STScI through grant JWST-AR-03252.001-A. J.S. is supported by a grant from the United States-Israel Binational Science Foundation (BSF), Jerusalem, Israel. N.N.S. is supported from NSF MPS-Ascend award ID 2212959. A.H. is supported by NSF award 2307788. This work was supported in part by the Swiss National Supercomputing Centre (CSCS) under allocations IDs s697, s698, and uzh18. C.K. would like to thank Camille Lemire for her whimsical comments.

The analysis and visualization of this work was accomplished using the following `Python` packages: `matplotlib` (J. D. Hunter 2007), `NumPy` (S. Van Der Walt et al. 2011), `SciPy` (P. Virtanen et al. 2020).

## Data Availability

The data supporting the plots within this article are available via doi:10.5281/zenodo.17993887. A public version of the GIZMO code is available at http://www.tapir.caltech.edu/~phopkins/Site/GIZMO.html. FIRE data releases are publicly available at http://flathub.flatironinstitute.org/fire.

## Appendix A
## Comparison of Three-dimensional Shapes

In Table A1, we provide a numerical breakdown for Figure 3. To understand the impact of our definition of 3D shape we directly compare the axis ratios of our three tensors to our quasi-observable 3D shape. In the top panel of Figure A1, we compare our $q_{\text{face-on}}$ to the equivalent $B/A$ measurement for All Stars (orange dotted line), Young Stars (blue dotted–dashed line), and *r*-band Luminosity-weighted Stars (red solid line). The lines represent the median values for $[B/A]/q_{\text{face-on}}$ for a given value of $q_{\text{face-on}}$ value, with the shaded region representing the $1\sigma$ scatter. For $q_{\text{face-on}} < 0.7$, all three tensor $[B/A]$ tend toward rounder shapes in the given orientation than $q_{\text{face-on}}$. Above that value, $q_{\text{face-on}}$ becomes rounder than Young Stars and Luminosity-weighted Stars.

The bottom panel of Figure A1 is the same as the top panel, but now we look at $[C/A]/q_{\text{edge-on}}$ for a given $q_{\text{edge-on}}$. While we again find All Stars to be more round, the Young Stars are always too elliptical compared to the mock observed $q_{\text{edge-on}}$. When looking at the relation of different tensor measurements, for both axis ratios All Stars > Luminosity-weighted Stars > Young Stars.

Following the conclusion from Figures 2 and 3, All Stars tend to be more spheroidal, and Young Stars have a lower $C/A$, which is consistent with more disky or elongated shapes. We determine that the Luminosity-weighted Stars is most consistent with our mock observed 3D shape and is therefore the most reliable comparison to the 3D shapes determined by observations. We do not expect the ratios to match between the *r*-band Luminosity-weighted Stars tensor and the PAO Sérsic measurements due to effects from different methods (tensor versus Sérsic fit) and from projection effects in the mock image such as extinction due to gas and dust.

**Table A1**
The Decomposition of Shapes Shown in Figure 3

| Measurement | Mass | Spheroid | Elongated | Disky | Triaxiality | Ellipticity |
|---|---|---|---|---|---|---|
| All Stars Tensor | $10^{8.5}$ | 0.98 | 0.02 | 0.00 | 0.55 | 0.39 |
| ... | $10^{9.5}$ | 0.99 | 0.00 | 0.01 | 0.40 | 0.46 |
| ... | $10^{10.5}$ | 0.48 | 0.00 | 0.52 | 0.16 | 0.62 |
| Young Stars Tensor | $10^{8.5}$ | 0.82 | 0.16 | 0.02 | 0.62 | 0.49 |
| ... | $10^{9.5}$ | 0.25 | 0.45 | 0.30 | 0.66 | 0.71 |
| ... | $10^{10.5}$ | 0.03 | 0.08 | 0.89 | 0.29 | 0.88 |
| Luminosity-weighted Stars Tensor | $10^{8.5}$ | 0.81 | 0.19 | 0.00 | 0.68 | 0.50 |
| ... | $10^{9.5}$ | 0.62 | 0.19 | 0.20 | 0.56 | 0.61 |
| ... | $10^{10.5}$ | 0.14 | 0.02 | 0.84 | 0.25 | 0.76 |
| *r* band | $10^{8.5}$ | 0.71 | 0.26 | 0.03 | 0.70 | 0.54 |
| PAO | $10^{9.5}$ | 0.63 | 0.24 | 0.13 | 0.59 | 0.60 |
| ... | $10^{10.5}$ | 0.37 | 0.07 | 0.56 | 0.37 | 0.67 |

**Note.** Measurement indicates the method used to measure the 3D axes, and mass is the given mass bin. Spheroidal, elongated, and disky are the fractional composition of each morphology given the measurement and mass bin. The triaxiality and ellipticity are the median values for the given sample.

(This table is available in its entirety in machine-readable form in the online article.)

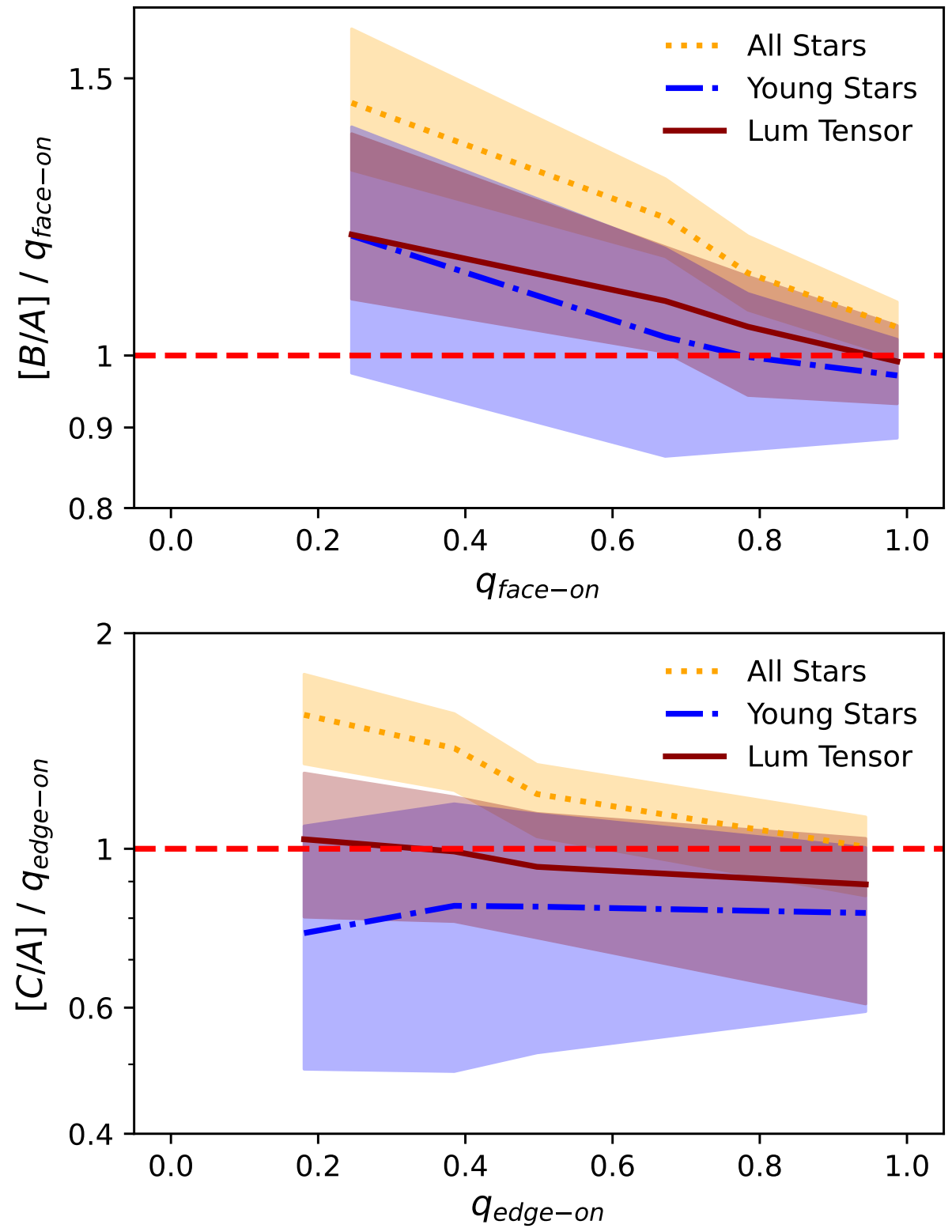

**Figure A1.** We compare the 3D axis ratios of the tensor measurements to the equivalent Sérsic axis ratios from the PAO images. The top panel shows the $B/A$ axis ratio for All Stars (orange dotted line), Young Stars (blue dotted–dashed line), and *r*-band Luminosity-weighted Stars (red solid line), divided by $q_{\text{face-on}}$, for a given $q_{\text{face-on}}$. The bottom panel is the same, but for $C/A$ and $q_{\text{edge-on}}$. The colored lines represent the median values, and the band represents the $1\sigma$ scatter.

## Appendix B
## Impact of Isophote Measurement on Morphologies

We explore a second method to characterize the 2D shape of each projected galaxy by measuring the ellipticity of the isophote that encloses 50% (or 90%) of the luminosity of the galaxy. Often the morphology of a specific isophote is investigated, however using the isophote that encloses 50% (or 90%) of the galaxy's light allows us to investigate the morphologies for a wide mass range of galaxies. Our goal is to test whether the shapes of our galaxies change systematically with decreasing surface brightness and if the isophote measurement differs from that derived by the Sérsic fit.

We fit ellipses to 100 isophote contours for each galaxy using a least squares method that minimizes the distance from each point along the contour to the fit ellipse. We exclude nonphysical fits that occurred when the best-fit ellipse has a very large radius and the contour data are a cluster of points in one small section of the ellipse. We identify the isophote corresponding to the ellipse that contains 50% (or 90%) of the galaxy light and define the morphology as the axis ratio for the given ellipse. We use the fits of isophotes within 15% of the 50% (or 90%) isophote to measure the variance in axis ratio. Similar to the issue discussed in Section 2.4, several of the 50% isophotes measured only a single burst of star formation. We did not use these fits in our analysis.

For all mass bins, the 50% isophote axis ratio distribution is shifted to slightly lower $q$ values compared to the 90% isophote. This implies that as you go to larger radii, galaxies tend toward more spherical morphologies. When compared to the Sérsic measurement, the 50% isophote $q$ distribution is shifted to slightly lower values for the low- and intermediate-mass bins. However in the high-mass bin, the isophote method is unable to produce axis ratios less than 0.24, and for all mass bins, the isophote method is unable to produce any $q < 0.2$. Therefore, this measurement method is not able to address the lack of low-mass disks and is not a good comparison to Sérsic measurements.

## Appendix C
## Impact of Observing Band on Morphologies

We show the axis ratio distributions measured from our randomly oriented images in the *u*, *g*, and *r* bands. As stated in Section 2.4, there are nonphysical fits in all three band, with up to 9% of the measurements being excluded in the *u* band. In Figure C1 we compare the axis ratio measurements for mock images with physical fits in all three bands. For the low- (left panel) and intermediate-mass (right panel) bins we show the axis ratio distribution for the *r* (solid red), *g* (dashed green), and *u* bands (dotted blue). The *u*-band distribution has a larger population of $q < 0.4$ measurements. This is consistent with the $u$ band tracking Young Stars, which tend toward more disky and elongated morphologies when compared with the $r$ band. This result stresses the importance of using measurements made in the same observing band for accurate comparisons.

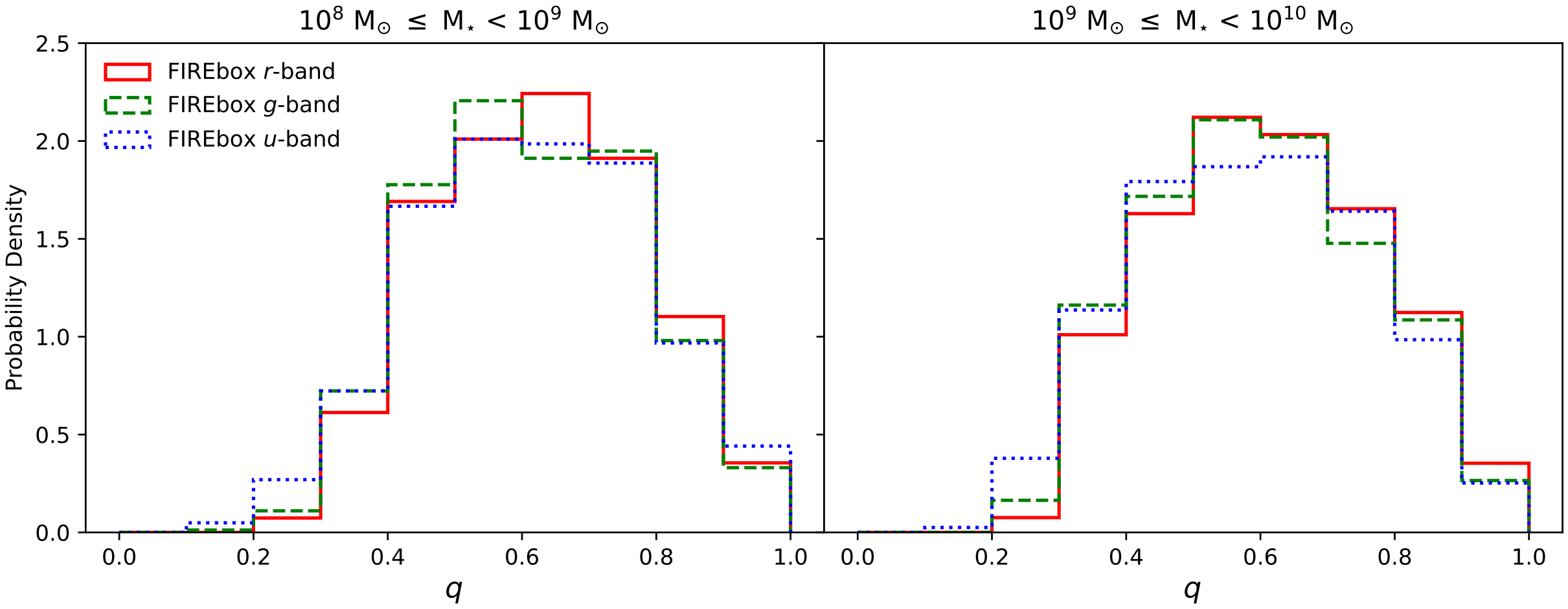

**Figure C1.** Similar to Figure 5, the 2D axis ratio distributions for FIREbox galaxies measure with a Sérsic fit in the $r$ (red solid line), $g$ (green dashed line), and $u$ bands (blue dotted line). For this figure, we include only mock images with well-fit $q$ values in all three bands.

## Appendix D
## Impact of Dust and SKIRT Simulation Parameters on Morphologies

In this appendix we look at a single, disky galaxy in FIREbox and explore how its inferred 2D $r$-band shape along different orientations depends on the dust modeling options in the radiative transfer code SKIRT. Our goal is to determine whether built-in SKIRT parameters such as wavelength bias, forced scattering, and dust-to-metal ratio may lead to significant deviations in the 3D and 2D shape distributions.

Wavelength bias is a number between zero and one that specifies the degree to which the wavelength spectrum is sampled from a custom wavelength distribution as opposed to the SED family (in this case, Bruzual and Charlot; P. Camps & M. Baes 2020). Most of the time, a higher wavelength bias produces a more evenly sampled spectrum by favoring low-luminosity areas in the wavelength range that may have been overlooked. On the other hand, a wavelength bias of zero guarantees that all wavelengths are sampled from the SED (classified as "Purely SED" in Table D1). This may affect our $r$-band images and subsequently the 3D axis ratios and 2D axis ratio distributions derived from their Sérsic fits. We find that altering the wavelength bias from 0.5 to 0.0 for a disky galaxy in the $10^{10}$–$10^{11}$ $M_\odot$ mass range does not significantly impact $q$ (Table D1). Thus, we apply a constant wavelength bias of 0.5 across our mock images.

Second, forced scattering in SKIRT allows the user to specify the minimum number of times that an emitted photon is scattered by dust. Because of the high optical depth of edge-on projections in disky galaxies, setting this option to true produces rather artificial-looking images. Furthermore, upon measurement of the same disky galaxy, we find that this parameter does not significantly impact its $q$. For this reason, we choose to set forced scattering to false.

**Table D1**
Comparison of $q$ Values for a Singular Disky Galaxy with Varying SKIRT Parameters

| Purely SED | Forced Scattering | High Dust Fraction | $q_{xy}$ | $q_{xz}$ | $q_{yz}$ |
| --- | --- | --- | --- | --- | --- |
| | | | 0.866 | 0.223 | 0.407 |
| | | ✓ | 0.823 | 0.265 | 0.392 |
| | ✓ | | 0.869 | 0.224 | 0.406 |
| | ✓ | ✓ | 0.820 | 0.265 | 0.391 |
| ✓ | | | 0.872 | 0.223 | 0.408 |
| ✓ | | ✓ | 0.820 | 0.265 | 0.391 |
| ✓ | ✓ | | 0.869 | 0.223 | 0.407 |
| ✓ | ✓ | ✓ | 0.819 | 0.265 | 0.390 |

**Note.** $q_{xy}$, $q_{xz}$, and $q_{yz}$ are the $q$ values derived from the Sérsic fit of the mock image projected along its three perpendicular orientations in the simulation box. The first row in the table corresponds to the $q$ values derived from mock images generated using the parameters in Section 2.3. The remaining rows show the new $q$ values for the mock images generated from an identical process except for the indicated change(s) in SKIRT parameter(s). Purely SED indicates a wavelength bias of zero; forced scattering indicates that forced scattering is set to "true;" and a high dust fraction indicates that $f_{\text{mass}} = 0.3$ as opposed to $f_{\text{mass}} = 0.1$. Notice that the largest changes in $q$ result from a increase in dust fraction.

We previously discussed the effects of varying $f_{\text{mass}}$, the dust-to-metal ratio, in Sections 2.3 and 3.2. To account for the higher gas-phase metallicities of high-mass galaxies in FIREbox, we lowered the dust-to-metal ratio to scale down the amount of dust that could have been extraneous from the high metallicities. We found that decreasing $f_{\text{mass}}$ across high-mass galaxies significantly increases the fraction of disks. A more robust dust model may help in resolving this tension and extrapolate the true fraction of disky galaxies in FIREbox. The combined effects of wavelength bias, forced scattering, and dust-to-metal ratio are summarized in Table D1.

## ORCID iDs

Courtney Klein https://orcid.org/0000-0002-2762-4046
Jenny D. Wang https://orcid.org/0009-0001-8916-8322
Luke Y. Xia https://orcid.org/0009-0002-4568-2423
James S. Bullock https://orcid.org/0000-0003-4298-5082
Jorge Moreno https://orcid.org/0000-0002-3430-3232
Robert Feldmann https://orcid.org/0000-0002-1109-1919
Francisco J. Mercado https://orcid.org/0000-0002-5908-737X
Claude-André Faucher-Giguère https://orcid.org/0000-0002-4900-6628
Jonathan Stern https://orcid.org/0000-0002-7541-9565
N. Nicole Sanchez https://orcid.org/0000-0001-7589-6188
Abdelaziz Hussein https://orcid.org/0000-0002-5560-8668
Michelle S. Park https://orcid.org/0009-0009-3348-6688